# Enhancing Graphene-derived Materials through Multimodal and Self-healable Interfaces

Yilun Liu

International Center for Applied Mechanics, SV Lab, School of Aerospace, Xi'an Jiaotong University, Xi'an 710049, China

Email: yilunster@gmail.com

Zhiping Xu

Department of Engineering Mechanics and Center for Nano and Micro Mechanics, Tsinghua University, Beijing 100084, China

Email: xuzp@tsinghua.edu.cn

**Recent studies have shown that graphene-derived materials not only feature outstandingly multifunctional properties, but also act as model materials to implant nanoscale structural engineering insights into their macroscopic performance optimization. Functionalizing the interfaces between graphene sheets by interlayer crosslinks has been proven to be an effective route to tune the mechanical properties. Here we explore the graphene-derived material with a layer-by-layer structure and multiple crosslinking mechanisms. The effects of multimodal and self-healable crosslinks are assessed in terms of interlayer loading transfer capability. The results show that the brick-and-motar hierarchy and synergetic effects from different crosslinks enable synergetic enhancement in the strength and toughness. The findings here could shed light on the development of high-performance paper-, fiber- or film-like macroscopic materials from monolayer nanosheets with nanoengineerable interfaces.**

## 1. Introduction

The synergetic excellence of mechanical, thermal and electronic properties of graphene has attracted immersed interests in, but not limited to, the materials community.[1] Focus has not only been placed on utilization of its high intrinsic strength up to 120 GPa and ultimate strain to failure of 20%, but also how to transfer the strong and tough performance of monolayer graphene nanosheets into macroscopic applications.[2] Within this scenario, nanostructures such as graphene, graphite nanoplatelets and carbon nanotubes have been widely utilized as reinforcing phases in high-performance composites.[3, 4] Major advantages following this approach include well-enhanced stiffness and resilience, thermal and electrical conductivities, and multifunctionality. However, it is also widely recognized that the native interface between these nanostructures and matrices without modification creates the weakest point in the mechanical sense. This critically prevents successful transfer of the ultrahigh strength in graphene across multiple length scales up to the macroscopic level and limits the toughness, in addition to other difficulties such as lacking of efficient technique to uniformly disperse nanostructures into the matrix at a high volume fraction.[4, 5]

Graphene-derived materials (papers, fibers, films etc.), in contrast to particle or fiber-based reinforce composites, feature many advantages, including not only extreme exposure of surface to the environment for functionalization, as well as rich and tunable crosslinking mechanisms between graphene sheets.[2, 6-8] Recently, a number of theoretical and experimental efforts have been made in predicting and optimizing mechanical performance of paper materials by taking the advantage of structural hierarchy that broadly appears in biological materials, such as bones, teeth and nacre [9, 10], where brittle minerals and soft proteins are integrated for their superior strength and toughness.[11] This is achieved by the staggered arrangement of the mineral platelets that distribute the tensile load, while the tensile load between neighboring mineral platelets is transferred by the shear in the soft protein interphase that can be well captured by a tension-shear chain model.[12] In this manner the high stiffness, strength of the mineral platelet and high toughness of protein both are utilized together to yield superior mechanical properties. Further works have been focused on the topological optimization and stabilization of the staggered layer-by-layer structures.[10] The graphene-derived paper or fiber material also feature a layer-by-layer structure.[3, 8, 13] Besides of those elegant insights into rational design of high-performance biocomposite materials, the graphene sheets here can be further crosslinked by different types of interactions in the interlayer gallery, such as covalent, dative, ionic, hydrogen bonds and van der Waals interactions.[6, 14, 15] By further considering the deformation of graphene sheets, a deformable tension-shear (DTS) model was proposed to describe the mechanical properties of graphene-derived materials [2], which predicts that giant graphene oxide sheets could be used to build ultrastrong materials, which was established by experimentalists recently[8]. Beyond these points, recent progresses in functionalize graphene sheets through various surface chemical groups [2, 6, 8, 14-18], and improved understanding of crosslinks in complex materials, such as multimodality, sacrificial bonds and self-healing behaviors [19-21], further enable material design in a new dimension, through engineering the interfacial properties at the molecular level.

In this work, we set the focus on introducing the concepts of multimodal and self-healable crosslinks into the layer-by-layer structure to construct graphene-derived superior materials, and explore their mechanical and failure properties through a combined approach including theoretical analysis, atomistic simulations, and discussion on the experimental results.

**2. Characterization of failure modes**

The mechanical behaviors of graphene-derived materials with a layer-by-layer microstructure under tension can be described in a deformable tension-shear (DTS) model by considering both intralayer elasticity and interlayer crosslinks as a continuum [2]. The in-plane tensile load within neighboring graphene sheets is transferred by shear force through the interlayer crosslinks. Based on a representative volume element (RVE) approach (**Figure 1a**), the DTS model shows that with a tension force $F_0$ on the RVE, the tensile strain $\varepsilon$ along the in-plane tensile direction in graphene and shear strain $\gamma$ in crosslinks are

$$\varepsilon = \partial u_1 / \partial x = F_0 \left[ \frac{1}{2} - \frac{\cosh(x/l_0)}{2} + \frac{1+c}{2s} \sinh(x/l_0) \right] / D \tag{1a}$$

$$\gamma = \frac{u_1 - u_2}{h_0} = \frac{F_0 l_0}{h_0 D} \frac{(1+c)\cosh(x/l_0) - s\sinh(x/l_0)}{s} \tag{1b}$$

respectively, where $u_1$ and $u_2$ are the displacement fields in neighboring graphene sheet 1 and 2, $h_0$ is the interlayer distance, $s = \sinh(l/l_0)$, $c = \cosh(l/l_0)$, $l$ is the size of RVE or half length of the graphene sheet, and the position $x$ is measured from the open end of graphene sheet. A parameter $l_0 = (Dh_0/4G)^{1/2}$ characterizing the length scale of interlayer loading transfer is defined here through effective interlayer shear modulus $G$, $h_0$ and the intralayer tensile stiffness $D = Yh$, where $Y$ and $h$ are the Young's modulus and thickness of the graphene sheet. The interlayer shear strain localizes near the ends of the graphene sheets and the highly efficient interlayer loading transfer regions is limited within $\sim l_0$ from the ends, see **Figure 1c**. The tensile strain $\varepsilon$ in graphene sheet maximizes at the center of a graphene sheet, while $\gamma$ in the crosslinking interphase maximizes at both the ends and center of the graphene sheet according to the staggered arrangement of the graphene sheets, as shown in **Figure 1c** for a model with $l/l_0 = 10$.

Under tensile load, the material could fail in two distinct modes: fracture in the graphene sheet at a critical tensile strain $\varepsilon_{cr}$ (denoted as mode G), or breaking of the interlayer crosslink beyond a critical shear strain $\gamma_{cr}$ (mode I). The selection of failure mode is determined by both structural and mechanical properties of the sheets and crosslinks. As both tensile and shear strain maximize at the central part of the graphene sheet along the tensile direction, fracture would nucleate there. The strain-based failure criteria are thus

$$\varepsilon(l) = \frac{F_0}{D} \leq \varepsilon_{cr} \tag{2a}$$

$$\gamma(l) = \frac{F_0 l_0}{h_0 D} \frac{1+c}{s} \leq \gamma_{cr} \tag{2b}$$

By defining two controlling parameters $k_1 = l/l_0$ and $k_2 = \gamma_{cr}(4Gh_0/D)^{1/2}/\varepsilon_{cr}$, the diagram of failure mode is illustrated in **Figure 1d**, and tensile strength of the paper is

$$\sigma_s = \begin{cases} \dfrac{D\varepsilon_{cr}}{2h_0} & k_2 \geq \dfrac{\sinh k_1}{1+\cosh k_1} \\ \dfrac{sD\gamma_{cr}}{2l_0(1+c)} & k_2 \leq \dfrac{\sinh k_1}{1+\cosh k_1} \end{cases} \tag{3}$$

In failure mode G, the graphene sheet is fractured at the center and the entire tension-shear chain is thus broken. Crack nucleates across the materials transversely and therefore no tensile load can be borne further. While in mode I the crosslinks fail from the ends and center of the graphene sheets, as illustrated in **Figure 1b**. The remaining crosslinks, however, could still carry the loads and thus no catastrophic failure occurs. The damaged material with partially broken crosslinks can be considered as a paper with reduced overlapping area. Especially, when large-scale graphene or functionalized graphene sheets are fabricated, the enhancement in the intersheet overlap could provide significant load transfer capability during progressive failure of the interlayer crosslinks, which yields high toughness in turn. This effect is also observed widely in aforementioned biological materials such as nacre, bone and teeth, where the high toughness of the materials is mainly contributed by proteins - the interphase of biological materials that bears shear.[10, 12]

Graphene or graphene oxides can be functionalized to fabricate layer-by-layer structures (papers, fibers, films etc.) with various types of crosslinks. In contrast to covalent bonds that feature high strength and stiffness but could hardly be reformed, dative, ionic, hydrogen bonds or van der Waals interactions could establish an interlayer crosslinking phase with an additional benefit that they can regenerate at the new equilibrium position after failure. Before discussing the impact of this fact on the macroscopic performance of graphene-derived materials, we will firstly explore, at the molecular level, the self-healing behaviors of interlayer crosslinks and its influence in the intersheet load transfer via atomistic simulations.

**3. Self-healable interlayer crosslinks**

As discussed earlier, interlayer crosslinks play an important role in determining the mechanical performance of graphene-derived hierarchical materials where interfacial failure is critical for their macroscopic performance. Here we explore the mechanics and failure behavior of hydrogen bond (H-bond) crosslinked graphene oxide papers as an illustrative example.[6] Molecular dynamics (MD) simulations were performed with focus on their tensile strength and toughness. In order to capture the bond breaking and (re)forming dynamics of crosslinks, we used the reactive force field (ReaxFF) where chemical reaction and charge redistribution are included.[22] Our models (**Figure 2a**) consist of two graphene layers with length $l$ = 16 nm and width $w$ = 4 nm. Periodic boundary conditions were applied in the width direction only. Tensile loads on the system was applied by constraining the left end of the supporting (bottom) layer, and pulling the right end of the top layer at a constant speed of $v$ = 2 m/s along the length direction, as illustrated in **Figure 2a**. Two model materials were explored here in our simulations. One consists of pristine graphene sheets, randomly functionalized with epoxy and hydroxyl groups [23] (**Figure 2a and 2c**). The chemical composition of the graphene oxide is $n_C$:$n_O$:$n_H$ = 1:0.25:0.125, within the typical range characterized in experiments.[23, 24] $n$ is the number density of atoms. The equilibrium distance between graphene oxide sheets at room temperature is 0.56 nm in the absence of interstitial water, which is close to the value obtained in our previous first-principles calculations.[2] In a second model, 10% vacancies are further introduced by removing carbon atoms from the graphene sheet (**Figure 2b and 2e**).

The relaxed atomic structures of both models are shown in **Figure 2a and 2b**, indicating clearly wrinkles due to the presence of defects in the basal plane. In the first model, these wrinkled sheets are firstly flattened before the interlayer crosslinks start to break under tensile load. For graphene oxides, all interlayer crosslinks contribute to interlayer load transfer after the sheets being straightened, and before the tensile force reaches a maximum. There exists a plateau in the tensile force-displacement curve (**Figure 2d**), following a reduction in the force amplitude. This can be explained as follows. (1) Functional groups are randomly distributed in the graphene sheet, so the equilibrium lengths of interlayer H-bonds differ. Due to this multimodality of crosslinks, interlayer crosslinks will break progressively under shear between functionalized graphene sheets that transfers tensile load, and thus yield a plateau in the tensile force. (2) Moreover, the H-bonds could break and reform reversibly during the relative sliding between the functional graphene oxide sheets at the interface, and thus load transfer could still be maintained after breaking of H-bonds begins, offering a self-healable interface. The reforming rate of H-bonds, which is the key parameter defining the self-healing behavior, depends on the density and distribution of the epoxy and hydroxyl groups in the new position after failure, as well as the loading rate. Once the H-bond interface cannot be fully healed tensile force will decrease to a lower value.

For the second model with ~10% vacancies in the functionalized graphene sheet, the tensile force-displacement relation is similar (**Figure 2d**). However, the amplitude of wrinkles is higher and thus larger stretch is required to straighten the sheets. Afterwards, tensile force continues to increase, and exceed the value in the first model without vacancies in the graphene oxide sheet. This is attributed to the enhanced interaction between interlayer crosslinks at defective sites. During the sliding failure of interface, H-bonds also reforms after breaking.

According to the simulation results, the effective shear strengths $\tau$ of the interlayer crosslinks in these two models are obtained as 279 and 321 MPa respectively, and the shear resistances at the self-healing stage are 101 MPa and 140 MPa, which are defined as the average shear stress during the sliding failure (see **Figure 2d**). It should be noticed that this enhancement of shear performance is established by reducing in-plane tensile strength in the functionalized graphene sheet concurrently. Furthermore the density and reforming rate of H-bonds can be tuned by adding water molecules into the interlayer of the graphene oxide sheets. This is evidenced in earlier studies [25], the tensile load can be effectively transferred by the H-bond networks between the water molecules and the graphene oxide, where the reforming rate of H-bonds is improved due to the mobility of water molecules.

From these two sets of simulations, two key features of interfacial crosslinks are elucidated in defining the load transfer process, i.e. the multimodal distribution and self-healable behavior during interfacial failure. The effects of these two characteristics are now to be investigated in an analytical model, in order to predict the overall mechanical properties of graphene-derived materials.

**4. Graphene-derived layer-by-layer materials with multimodal and self-healable interfaces**

Graphene-derived materials such as graphene oxide papers, fibers and films feature hierarchical structures (the interlayer crosslinks, distribution, size and stacking of graphene sheets).[3, 7, 8] In our previous work [2], we found that the mechanical properties of graphene-derived papers can be finely tuned by adjusting the structure and distribution of graphene sheets and crosslinks. However, observations here suggest that introducing multimodal crosslinks can simultaneously enhances the stiffness, strength and toughness of graphene-derived materials. For example, bimodal crosslinks with both self-healable short crosslinks (SCs, e.g. dative, ionic, hydrogen bonds, and van der Waals interactions [6, 15]) and long crosslinks (LCs, e.g. covalent bonds through polymer intercalation [14]) could introduce effective strengthening and toughening mechanisms. In recent experiments, it is reported that both strength and stiffness of graphene oxide fibers are improved by introducing ions (e.g. calcium, magnesium, boron) into the gallery regions [8, 15, 16] that yields a record tensile strength ~0.5 GPa, and covalent crosslinking using glutaraldehyde or poly(vinyl alcohol) improves load-bearing capability of graphene-derived papers.[14, 18]

The mechanics of multimodal networks has been studied in rubberlike materials consisting of both long and short polymer chains in a crosslinked network.[19] Despite of the similarity in concept, synergetic enhancement by LCs and SCs could be established in graphene-derived materials effectively due to the layer-by-layer structure and outstanding mechanical properties of single sheet. Thus the stiffness, strength and toughness can be improved simultaneous due to synergistic effects of LCs and SCs. Several additional merits arise immediately by introducing the multimodal self-healable interlayer crosslinks: (1) self-healable SCs act as a sacrificial interface to continuously dissipate the mechanical energy by breaking and reconstruction of the crosslinks, while the LCs maintain the structural integrity, (2) the self-healable feature of the crosslinks enables robust performance for cyclic loading, (3) the SCs keep the compact layer-by-layer structure of the graphene-derived materials, which is critical for their high mechanical performance, (4) SCs could also prevent structural failure of graphene-derived papers under complicated load conditions such as bending, torsion, and moisture-induced swelling.

As illustrated in **Figure 3a**, a two-dimensional RVE of graphene paper was constructed by considering graphene sheets with a uniform lateral size $2l$, where adjacent layers are crosslinked by bimodal agents of covalent LCs and self-healable SCs. Here the mechanical response of the LCs was assumed to be hyperelastic. The constitutive relation between shear stress and shear strain is $\tau_L = G_L(\lambda - 1/\lambda^2)/3$. Here we assume the interlayer crosslinks to be pure shear. The longitudinal tensile deformation $\lambda$, i.e. the ratio of the extended length to the length of LCs at rest, is expressed in the interlayer shear strain $\gamma$ as $\lambda = 1 + \gamma$.[19] The critical failure shear strain of LCs is assumed to be $\gamma_L$. For deformation in the small strain regime the relation degenerates to the linear relation of $\tau_L = G_L \gamma$ and we define $G_L$ as the effective shear modulus of the LCs. The mechanical responses of SCs are simplified into a linear relation with self-healable feature, where the interlayer shear stress contributed by SCs is $\tau_S = G_S\gamma_s$. $G_I$ is the effective shear modulus of SCs and $\gamma_s$ is the interlayer shear strain of SCs after their last reconstruction, which is defined as $\gamma_s = u_1(x_{-1}, t_{-1}\text{->}t) - u_2(x_{-1}, t_{-1}\text{->}t)]/h_0$. Here $t_{-1}$ and $x_{-1}$ are the time and position of SCs at the last reconstruction, and $u_{1(2)}(x_{-1}, t_{-1}\text{->}t)$ is the displacement of graphene sheet 1(2) from time $t_{-1}$ to $t$ at the point $x_{-1}$ (**Figure 3**). The critical failure shear strain of the SCs was assumed as $\gamma_I$. The reconstruction of SCs is a rate-dependent process. However in this work time $t$ only indicates the load history and the effect of loading rate will be explored in our future investigation.

By including combined effects of LCs and SCs, the constitutive relation of interlayer crosslinks is schematically shown in **Figure 3b**. As the shear strain increases beyond $\gamma_I$, SCs fails so the shear stress drops down. After the reconstruction at a new equilibrium position, the shear stress gradually increases again as the shear deformation further proceeds. During the repeating breaking and reconstructing processes of the SCs, mechanical energy $E_D = nG_I\gamma_I^2$ corresponding to the shaded area in **Figure 3b** is dissipated, which depends on the number of self-healed SCs $n$, their effective shear modulus and strain amplitude.

The mechanical properties of graphene derived layer-by-layer materials with multimodal and self-healable interlayer crosslinks can be estimated by the DTS model in a RVE approach. According to our previous study [2], we assumed that the interlayer crosslinks are uniformly distributed. By substituting the expression of shear stress into the DTS model [2] we obtained the governing equations

$$D\frac{\partial^2 u_1(x,t)}{\partial x^2} = 2(\tau_L + \tau_S) \tag{4a}$$

$$D\frac{\partial^2 u_2(x,t)}{\partial x^2} = -2(\tau_L + \tau_S) \tag{4b}$$

As SCs are usually stiffer and more brittle than LCs (where entropic elasticity governs), here we assume $G_S = 10G_L$ and $\gamma_S = 0.1\gamma_L$ in the following discussion. The length of graphene sheet is $2l = 4l_0$, where $l_0 = (Dh_0/4G_L)^{1/2}$. Here we consider three representative systems with (1) both LCs and self-healable SCs, (2) only LCs, and (3) only self-healable SCs. In the following analysis, we will focus on the deformation range where LCs are not broken, and the maximum shear strain for the material is determined by the critical failure shear strain of LCs $\gamma_L$. We also assume that SCs are immediately reconstructed at the new position.

The stress-strain relations for these three model systems are predicted by solving Eq. (4) numerically using the conjugate gradient method. The results are summarized in **Figure 4**. According to the combined effects from LCs and SCs, virtues such as the high stiffness and self-healability of SCs, high extension and toughness of LCs are fully utilized and transferred into the macroscopic material properties for *system 1*. The stiffness in the small deformation regime is thus close to that of SCs, and the ultimate tensile strain is even larger than that of LCs (see the dash and dash-dot lines in **Figure 4a**). This is consistent with the experimental results that the ultimate tensile strain can be significantly improved by adding the water molecules into glutaraldehyde-crosslinked graphene oxide papers.[14] Due to the self-healability of SCs, mechanical energy is dissipated by rupturing and reconstruction of SCs, which is indicated by the hysteresis between loading and unloading curves as shown in **Figure 4a**. It should be noticed that during the unloading process, the deformation of graphene-derived materials - the rupturing and reconstruction of SCs - is not elastic or reversible.

In contrast, for the interlayer crosslinks with SCs only (*system 3*), the interlayer load transfer is slightly enhanced after reconstruction of SCs (see the nonlinear region in **Figure 4a**) and the strength is ~45% of the value in *system 1*. There is also hysteresis observed within the loading and unloading cycles, and the dissipated energy in one cycle is ~70% of *system 1*. However, one should be noticed that due to the absence of LCs, the integrity of the layer-by-layer structure could be destroyed during rupturing events of SCs, especially when complicated loading conditions, such as bending and torsion, are applied.

A diagram for the tensile strength of the graphene-derived materials with bimodal crosslinks (both LCs and self-healable SCs) normalized by the strength of graphene paper with LCs only is plotted in **Figure 4b**. Here the length of the graphene oxide sheet is $4l_0$, and SCs with parameters of $G_S/G_L$ (from 2 to 20) and $\gamma_L/\gamma_S$ (from 2 to 20) are considered. The strength improvement increases as $G_S$ and $\gamma_S$ increase. Based these results the strength enhancement by including self-healable SCs can be described through a parameter $G_S\gamma_S/G_L\gamma_L$ (**Figure 4c**). Both parameters $G_S$ and $\gamma_S$ of SCs contribute to the enhancement. $G_S$ improves the stiffness of graphene paper, and as $G_S$ is usually much larger than $G_L$, the stiffness of model *system 1* is close to the stiffness of model *system 3* (**Figure 4a**). While as the rupturing of SCs leads to kinks in the tensile force curves, lower value of $\gamma_I$ could yield smoother response in the force and more stabilized performance of graphene-derived materials. As the rupturing and reconstructing events of SCs increase, the energy dissipation also increases.

Although our results in **Figure 4c** show that the strength enhancement of graphene-derived materials decreases as the size of the single sheet increases, the absolute value of the tensile strength still increases. According to the results obtained from the DTS model, the enhancement by enlarging the sheet size almost converges as the $l > 8l_0$. In our study, we focus on the regime before LCs are broken. Indeed, the failure of LCs can also progressively develop from the ends of the graphene sheet to the center (mode I), meanwhile the SCs are repeatedly ruptured and reconstructed, so the energy dissipation and toughness could be further enhanced for sheet of large sizes.

According to our previous analysis, the interlayer shear strain localizes at the edges of the graphene sheets for relatively large-size graphene sheet (measured by $\sim l_0$ from the edges), where the failure and reconstruction of SCs occur as well. The tensile load in the graphene sheet is transferred by interlayer shear load, where both LCs and SCs contribute.

Because of the failure and reconstruction of SCs, we can assume that the average shear stress (over different states from the newly reconstruction to nearly failure) contributed by SCs is $G_S\gamma_S/2$ within $l_0$ from the edges. For each graphene sheet, the effective load transfer length of SCs is thus $4l_0$ by considering its two sides and two edges (see the illustration of RVE in **Figure 3a**). As a result the tensile force transferred by the interlayer shear of SCs is $2G_S\gamma_S l_0$ and the effective tensile stress is $2G_S\gamma_S l_0/2h_0$. Based on our previous work, the tensile stress of only LCs is then

$$\sigma_L = \frac{s\gamma_L D}{2(1+c)l_0} \quad (5),$$

and the enhancement by LCs and SCs is

$$\sigma_M/\sigma_L = 1 + \frac{1+c}{2s}\frac{G_S\gamma_S}{G_L\gamma_L} \quad (6),$$

where $l_0 = (Dh_0/4G_L)^{1/2}$. The prediction from **Eq. (6)** works well for relatively large-size graphene sheet and small values of $G_S\gamma_S/G_L\gamma_L$ (**Figure 4c**). This is because that the shear strain localizes near the edges in large-size graphene sheets and for relative small $G_S\gamma_S/G_L\gamma_L$ the contribution of SCs to the shear strain is insignificant.

The stiffness, strength and toughness of graphene-derived materials can be simultaneously enhanced by introducing LCs and self-healable SCs. The strength of LCs plus SCs is about the sum of their individual performance. Besides, the ultimate tensile strain can be higher than the system with LCs only. If only LCs are available, the stiffness and strength are usually low. This could be improved by increasing the density of LCs that is equivalent to increase the effective shear modulus $G_C$ of LCs.[18] However, the density of LCs is limited by the number density of anchoring sites in the graphene sheets and the layer-by-layer microstructure (one LCs may entangle with others at a high density). On the other hand, if there are only non-self-healable SCs at the interface (see the linear region in **Figure 4a**), the strength and toughness is usually low due to the limited deformability of SCs. The self-healing SCs can improve partly the toughness and slightly the strength, but the layer-by-layer structure could become unstable during rupturing of SCs.[8, 16] Moreover, the discussion above is based on the failure mode I. If the failure is in mode G then the toughening mechanism by interlayer crosslinks will not work.

## 5. Conclusion

In this study we explore the graphene-derived materials by introducing multimodal self-healable interlayer crosslinks into a layer-by-layer structure. The role of multimodal and self-healable crosslinks in enhancing mechanical properties of graphene papers is elucidated through atomistic simulations and theoretical analysis. Nanoscale confinement between functionalized graphene sheets enables efficient load transfer that is critically defined by the crosslinks. Optimized hierarchical materials consisting of large-size graphene sheets and multimodal crosslinks that include both long, strong and short self-healable ones offer the opportunity to make use of the outstanding mechanical properties of graphene to the largest extend. The models and results here provide key concepts in optimal design of graphene materials in their macroscopic forms. The rationales explored here can be readily extended to other materials made of low-dimensional nanostructures that could be functionalized, such as boron nitride and molybdenum disulfide sheets.

**Acknowledgments**

This work was supported by the National Natural Science Foundation of China through Grant 11222217, 11002079, Tsinghua University Initiative Scientific Research Program 2011Z02174, and the Tsinghua National Laboratory for Information Science and Technology of China.

**FIGURES AND CAPTIONS**

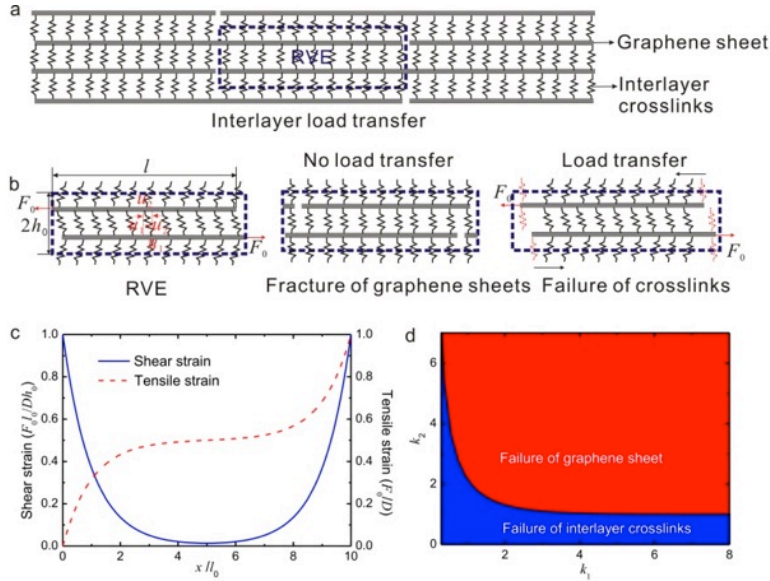

**Figure 1.** (a) The representative volume element (RVE) of the deformable tensile-shear (DTS) model including two graphene sheets with an overlap at half of its length. (b) The two failure modes as explained in the text - failure of the sheets or the crosslinks. (c) The distribution of tensile and shear strain for $l/l_0 = 10$. (d) Failure modes of graphene papers that is tuned by two parameters $k_1$ and $k_2$.

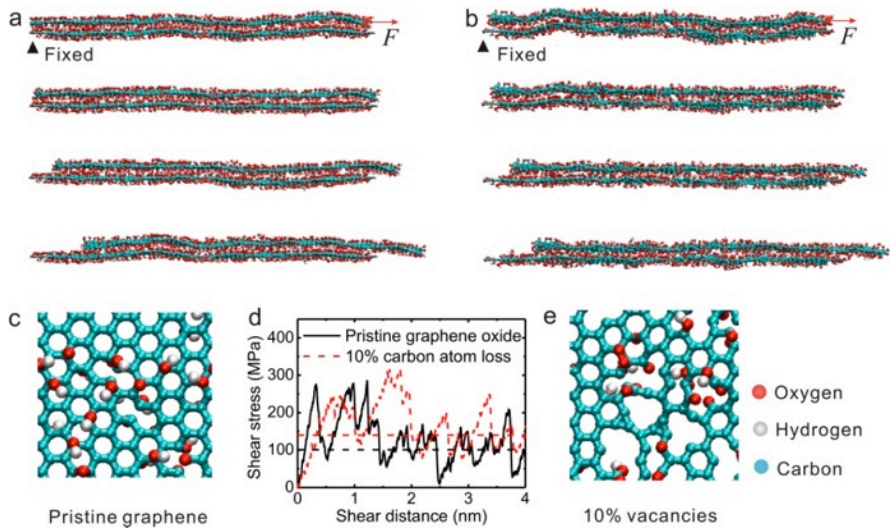

**Figure 2.** Failure of crosslinks between neighboring graphene oxide sheets with (a) and without (b) basal-plane vacancy-type defects. (c) and (e) The detailed structures of graphene oxide sheets corresponding to (a) and (b). (d) Relationship between the tensile force applied at ends of loaded sheet and the end displacement. The dashed lines indicate the average shear resistance during sliding failure for pristine graphene oxide, and graphene oxide with 10% vacancies in the basal plane.

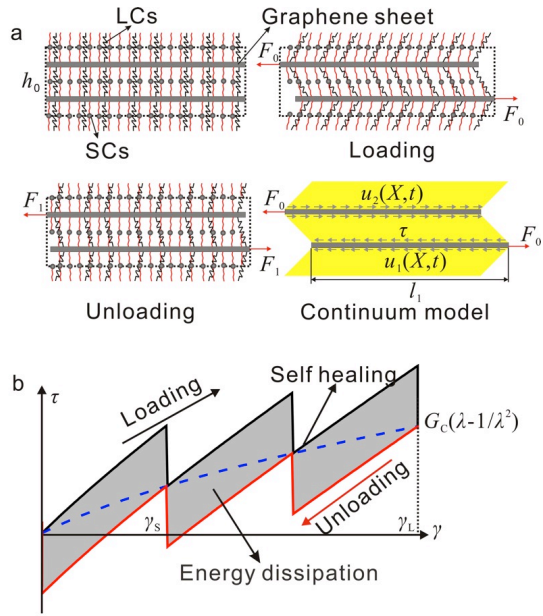

**Figure 3.** Graphene papers as crosslinked bimodally in a layer-by-layer structure. (a) An illustration of the model and mechanics of graphene papers with bimodal interlayer crosslinks. Red and black springs represent for short crosslinks (SCs) and long crosslinks (LCs) between graphene sheets (gray thick lines). $F_0$ is the tensile force applied to the unit cell, under which LCs start to fail from the center of graphene sheets where shear strain maximizes, then the SCs breaks and reconstructs. (b) Shear stress-strain relationship by considering the bimodal crosslinking and self-healing mechanisms.

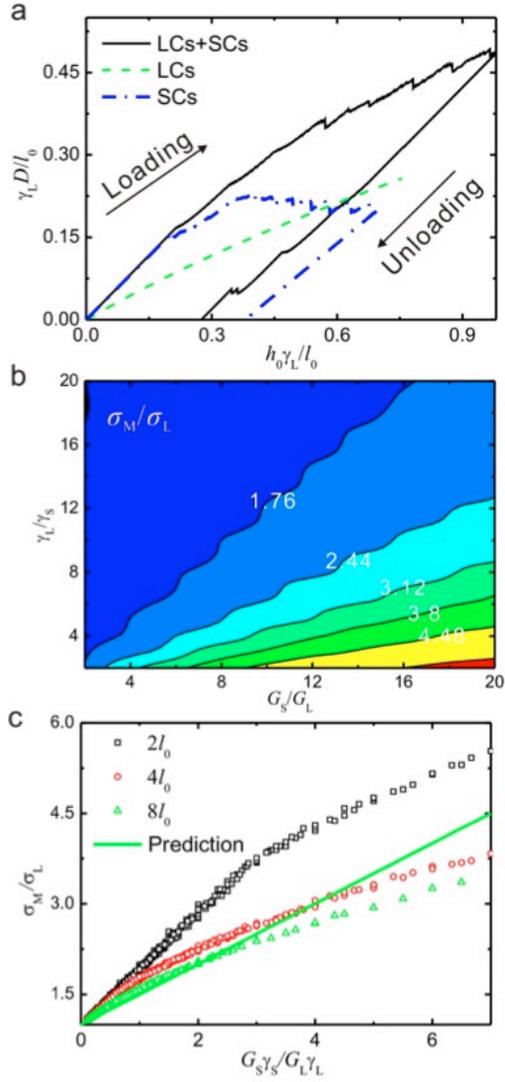

**Figure 4.** (a) The stress-strain relation for the three types of interlayer crosslinks. With LCs only, the stress-strain relations are elastic and no hysteresis between loading and unloading curves. (b) The strength enhancement of graphene paper $\sigma_M/\sigma_L$ with bimodal crosslinks and self-healable interfaces for different $G_S$ and $\gamma_S$, where $\sigma_L$ is strength with LCs only. (c) The relation between the strength enhancement and $G_S\gamma_S/G_L\gamma_L$ for three different length of graphene oxide sheets, $l = 2l_0$ (squares), $l = 4l_0$ (circles) and $l = 8l_0$ (triangles). The solid line is theoretical prediction from **Eq. (6)**.